\documentclass[5p]{elsarticle}
\usepackage{graphicx,subfigure}
\usepackage{amsfonts}
\usepackage{amsmath,amsthm,amssymb}

\begin{document}
\title{Bifurcations in the optimal elastic foundation for a buckling column}

\author[Not]{Daniel Rayneau-Kirkhope}
\author[Unilever,Maths]{Robert Farr} \author[Fudan]{K. Ding} 
\author[Not]{Yong Mao}
\address[Not]{ School of Physics and Astronomy, University of
  Nottingham, Nottingham, NG7 2RD, UK }
\address[Unilever]{Unilever R\&D, Olivier van Noortlaan 120, AT3133, Vlaardingen, The Netherlands }
\address[Maths]{London Institute for Mathematical Sciences, 22 South Audley Street, Mayfair, London, UK}
\address[Fudan]{Department of Physics, Fudan University,
Shanghai, 200433,
People's Republic of China}

\begin{abstract}
We investigate the buckling under compression of a slender beam with a 
distributed lateral elastic support, for which there is an associated cost. 
For a given cost, we study the optimal choice of support to protect against 
Euler buckling. We show that with only weak lateral support, the optimum distribution 
is a delta-function at the centre of the beam. When more support is allowed, 
we find numerically that the optimal distribution undergoes a series of bifurcations. 
We obtain analytical expressions for the buckling load around the first 
bifurcation point and corresponding expansions for the optimal position 
of support. Our theoretical predictions, including the critical exponent of 
the bifurcation, are confirmed by computer simulations.

\end{abstract}


\maketitle

\section{Introduction}

Buckling is a common mode of mechanical failure \cite{Timoshenko}, and its 
prevention is key to any successful engineering design. 
As early as 1759, Euler \cite{Euler}
gave an elegant description of the buckling of a simple beam, from which
the so-called Euler buckling limit was derived. 
Works which cite the goal of obtaining structures of least weight stable against buckling can be 
found throughout the literature \cite{Lagrange,SJCox,Weaver,Budiansky,Tian}, and much
understanding has been gained on optimal structural design 
\cite{Timoshenko,Cox,Gordon}. 
Designs of ever increasing complexity have been analysed and recent work suggested that the optimal 
design of non-axisymmetric columns may involve fractal geometries 
\cite{Farr1,FarrYong}.
With the development of powerful computers,
more and more complicated structures can be designed with optimised mechanical
efficiency. However, understanding and preventing buckling remains as relevant
as ever.

\begin{figure}
\begin{center}
\subfigure[]{\includegraphics[width=3.0in]{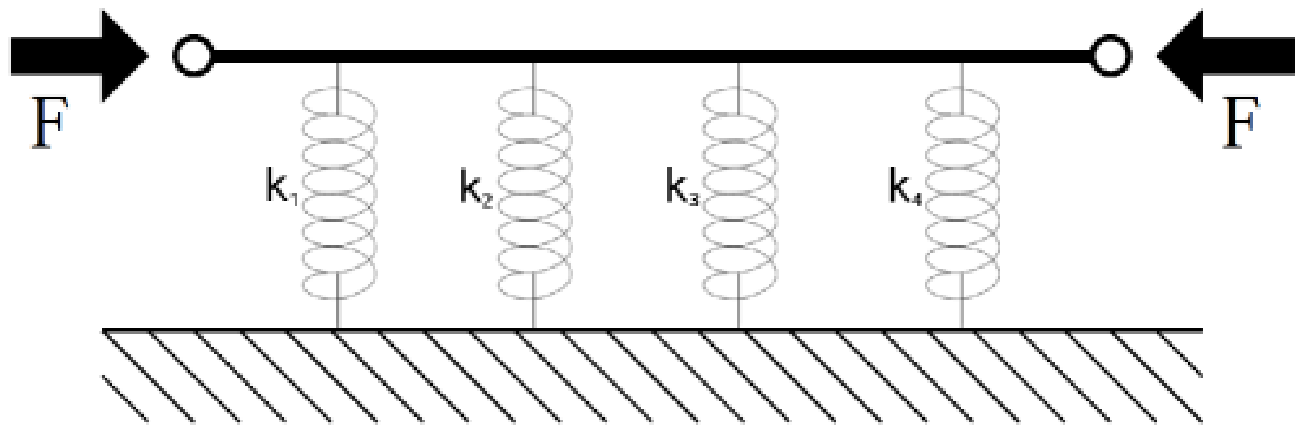}\label{springs}
}
\subfigure[]{\includegraphics[width=3.0in]{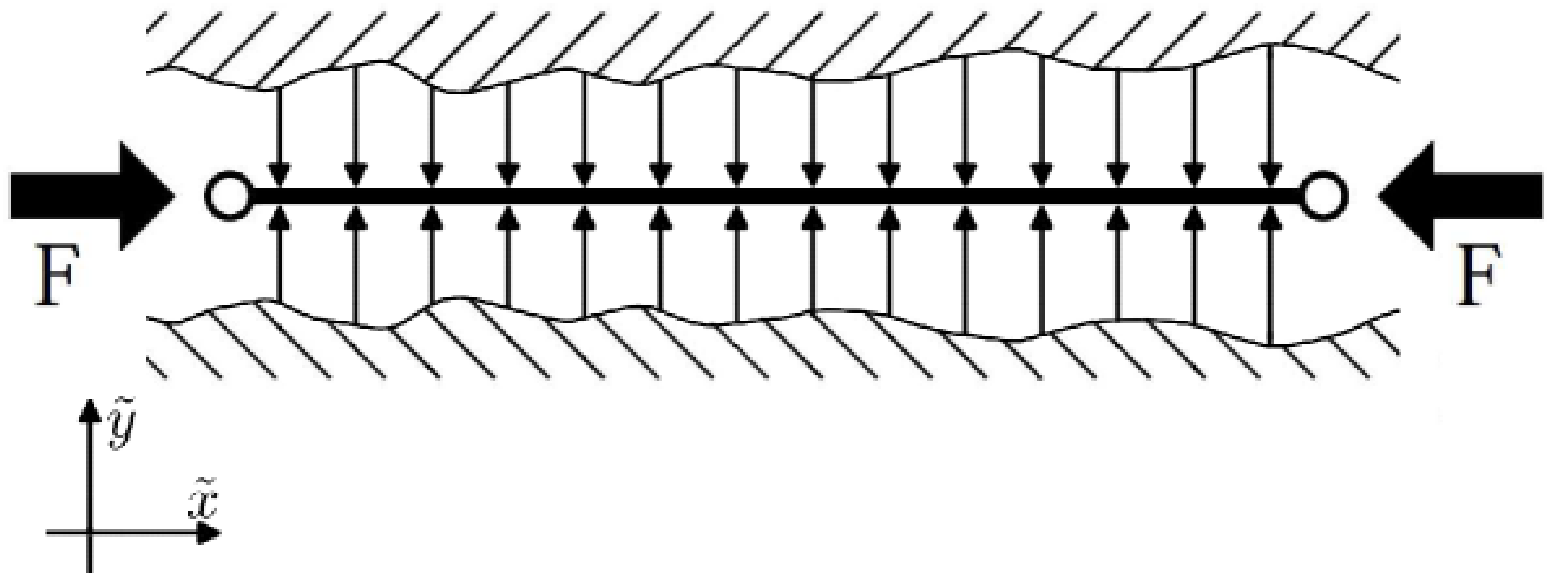}
\label{cont}}
\caption{\label{beam1}
Schematic of a slender beam with elastic support, loaded under
compression force $F$. (a) shows the case where the lateral restoring
force per unit length along the beam $q(\tilde{x})$ is provided by
linear springs of spring constant $\{k_{i}\}$ at discrete points
$\{\tilde{x}_{i}\}$, so that
$q(\tilde{x})=\sum_{i=1}^{4}k_{i}\delta(\tilde{x}-\tilde{x}_{i})\tilde{y
}/L$. (b) shows schematically the case where there is a continuous
support: the lengths of the arrows indicate the local spring constant.
}
\end{center}
\end{figure}

In this paper, we consider a simple uniform elastic beam, freely hinged
at its ends and subjected to a compressive force and therefore
vulnerable to buckling. However, in contrast to Euler's original
problem, we specify that the beam is stabilized by restoring forces,
perpendicular to its length, which are provided by an elastic foundation
(as illustrated in figure \ref{beam1}). This represents a simple and practical
method of protecting against buckling instabilities.

In the simplest case figure \ref{springs}, we can imagine this elastic foundation
as a finite collection of linear springs at points along the beam. Each
has a spring constant, and so provides a restoring force at this point,
proportional to the lateral deflection of the beam. More generally, the
elastic foundation could be distributed as a continuous function along
the length of the beam, rather than being concentrated into discrete
springs (figure \ref{cont}). In this case, there is a spring constant per unit
length, which may vary along the beam.

We are interested in optimising this elastic support, and so we need to
specify a cost function for it. This we take to be the sum of the spring
constants (if there are a discrete collection of springs) or the
integral over the spring constant per unit length along the beam (if the
elastic foundation is continuous). By choosing the optimal distribution
of these spring constants, we wish to find the minimum cost of elastic
support which will protect against buckling under a given compressive
load (or equivalently, the distribution of an elastic support of fixed
cost which will support the maximum force).

The optimal position of one or two deformable or infinitely stiff
supports have been studied in the literature (see for example Ref.
\cite{Olhoff} and references therein), and general numerical approaches
established for larger numbers of supports \cite{Olhoff}. However, in the
present paper, we consider the general case where any distribution of
support is in principle permitted.

A perturbation analysis shows that in the limit of weak support strength,
the optimal elastic foundation is a concentrated delta-function
 at the centre of the beam, but when stronger supports are permitted, we show that
the optimal solution has an upper bound on the proportion of the beam that remains unsupported. 
In this sense, the optimum distribution becomes more uniform for higher values
of support strength. To tackle the problem in more detail,
we develop a transfer matrix 
description for the supported beam, and we find numerically
that the optimal supports undergo a series of 
bifurcations, reminiscent of those encountered in iterated maps. However,
we are only able to proceed a limited distance in the parameter space
and we are unable to explore for more complex behaviour (for
example, any possible signature of chaos \cite{Chaos}).

We obtain analytic expressions for the buckling load in the vicinity
of the first bifurcation point and a corresponding series expansion for the 
optimal placement of elastic support. Following this optimization we show that a 
mathematical
analogy between the behaviour exhibited in this problem and that found in 
Landau theory of second order phase transitions\cite{Landau} exists.  
However, the analogue of free energy is non-analytic, while in Landau theory 
it is a smooth function of the order parameter and the control variable.
Our results,
including critical exponents are confirmed by computer simulations, and 
should provide a basis for future analysis on higher order bifurcations.

\section{Theory}

A slender beam of length $L$, hinged at its ends, under a compressive
force $F$, is governed by the Euler-Bernoulli beam
equation \cite{Timoshenko}:
\begin{equation}
EI\frac{d^4\tilde{y}}{d\tilde{x}^4} + F \; \frac{d^2 \tilde{y}}{d\tilde{x}^2} + q(\tilde{x}) =0,
\label{eq:ebu}
\end{equation}
where $E$ is the Young modulus of the beam, $I$ is the second moment of its
cross sectional area about the neutral plane, $\tilde{y}$ is the
lateral deflection, $\tilde{x}$ the distance along the beam and $q(\tilde{x})$
is the lateral force applied per unit length of beam. The beam is freely hinged at its end points and
therefore the deflection satisfies $\tilde{y}=\tilde{y}''=0$ at $\tilde{x}=0$ and $L$. 

If the lateral force is supplied by an elastic foundation, which
provides a restoring force proportional to the lateral deflection, then 
through rescaling we introduce the following 
non-dimensional variables 
$x=\pi\tilde{x}/L$, $y=\pi\tilde{y}/L$, $f=FL^2 /(EI\pi	^2)$ and
$\rho =qL^4/(EI\pi^4 \tilde{y})$.
Eq.\ (\ref{eq:ebu}) becomes
\begin{equation}
\frac{d^4y}{dx^4} + f \; \frac{d^2 y}{dx^2} + \rho(x)y =0
\ \ \ {\rm for}\ \ \ x\in (0,\pi),
\label{eq:eb}
\end{equation}
where $y(0)=y(\pi)=y''(0)=y''(\pi)=0$ and
$\rho(x)$ represents the strength of the lateral support (for example
the number of springs per unit length) at position $x$.

We are always interested in the minimum value $f_{\min}$ of $f$
that leads to buckling [in other words, the smallest eigenvalue
of Eq.\ (\ref{eq:eb})]. For the case of no support ($\rho=0$), the
possible solutions to Eq.\ (\ref{eq:eb}) are $f\in\mathbb{Z}^{+}$, and
so buckling first occurs when $f=1$.

Lateral support improves the stability (increasing the minimum value
of the applied force $f$ at which buckling first occurs), but
we imagine that this reinforcement also has a cost. In particular, for
a given value of
\begin{equation}
m\equiv \int \rho(x) \; dx ,
\end{equation}
we seek the optimal function $\rho(x)$ which maximises the minimum buckling
force $f_{\min}$.

The simplest choice we can imagine is that $\rho$ takes the uniform
value $m/\pi$, so that the form of deflection is $y(x)\propto\sin kx$,
for some integer $k$, which represents a wavenumber.

This leads immediately to the result that in this case
\begin{equation}
f_{\min}=\min_{k\in \mathbb{Z}}\left[ k^2 + \frac{m}{\pi k^2}\right].
\label{const}
\end{equation}

Eq.\ (\ref{const}) has a physical interpretation: the first term comes from
the free buckling of the column which is most unstable to buckling
on the longest allowed 
length scales (i.e. the smallest values of $k$), as demonstrated by
Euler. The second term represents the support provided by the elastic
foundation, which provides the least support at the shortest length scales
(largest values of $k$). 
The balance between these two terms means that as $m\rightarrow\infty$, 
the uniformly supported 
column buckles on a length scale of approximately 
\begin{equation}
l_{\rm eff}\approx 
(\pi/m)^{1/4}
\ \ \ {\rm as}\ \ \ m\rightarrow\infty, 
\label{leff}
\end{equation}
and can support a load
\begin{equation}\label{uni}
f_{\rm uni}\sim 2\sqrt{m/\pi}.
\end{equation}

Now, although a uniform elastic support is easy to analyse, it is clear that
this is not always optimal. Consider the case where $m$ is very small,
so that $\rho$ provides a small correction in Eq.\ (\ref{eq:eb}). In this
case, the eigenvalues remain well-separated, and we can treat the equation 
perturbatively: let
\begin{equation}
y=y_{0}\sin x + y_{1}(x)\ \ \ {\rm and}\ \ \ f=1+f_{1},
\end{equation}
then from Eq.\ (\ref{eq:eb}), if we multiply through by $\sin x$ (the
lowest unperturbed eigenfunction) and
integrate, we have to leading order:
\begin{equation}
\int_{0}^{\pi}\left\{
\sin x\left[\frac{d^4y_{1}}{dx^4} + \frac{d^2 y_{1}}{dx^2}\right] + 
y_{0}\sin^{2}x \left[\rho-f_{1}\right]
\right\} {\rm d}x=0.
\end{equation}
Repeated integrations by parts with the boundary conditions
$y_{1}''=0$ at $x=0,\pi$ establishes the self-adjointness of the
original operator, and we arrive at
\begin{equation}
f_{1}=\frac{2}{\pi}\int_{0}^{\pi}\rho(x)\sin^{2}x\; {\rm d}x.
\end{equation}

We therefore see that in the limit $m\rightarrow 0$, the optimal
elastic support is $\rho (x)=m\delta(x-\pi/2)$, and for this case,
$f_{\min}=1+(2m/\pi)+O(m^2)$.

The requirement for optimal support has therefore concentrated the elastic 
foundation into a single point, leaving the remainder of the beam unsupported.

\section{Transfer Matrix formulation}

In order to proceed to higher values of $m$ in the optimization problem, we assume
that there are $N-1$ discrete supports at the positions $\{x_{n}\}$, 
with corresponding set of scaled 
spring constants $\{\beta_{n}\}$, adding up to the total $m$:
\begin{equation}
\rho(x)=\sum_{n=1}^{N-1} \beta_n \delta(x-x_n)
\end{equation}
\begin{equation}
m=\sum_{n=1}^{N-1} \beta_n \;
\end{equation}
These discrete supports divide the beam into $N$ (not necessarily equal) 
segments, and for convenience in later calculations, we also define
the end points as $x_{0}\equiv 0$ and $x_{N}\equiv\pi$.

For each segment of the beam given by $x_{n}<x<x_{n+1}$, the Euler-Bernoulli
equation (\ref{eq:eb}) can be solved in the form 
\begin{eqnarray}
y(x)=A_n \sin[f^{1/2}(x-x_{n})]
+
B_n \cos[f^{1/2}(x-x_{n})] \nonumber \\
+
C_n (x-x_{n})+
D_n. \label{piece}
\end{eqnarray}
If we integrate Eq.\ (\ref{eq:eb}) over a small interval around $x_{n}$,
we find that,
\begin{eqnarray}
 y\left(x_n^+\right) = y\left(x_n^-\right), \quad y'\left(x_n^+\right) = y'\left(x_n^-\right), \nonumber \\
 y''\left(x_n^+\right) = y''\left(x_n^-\right), \nonumber \\
 y'''\left(x_n^+\right) - y'''\left(x_n^-\right) + \beta_n y\left(x_n\right) = 0,
\end{eqnarray}
where $x_n^+$ and $x_n^-$ are values infinitesimally greater and less than than $x_n$ respectively.
Defining $\mathbf{v}_n\equiv(A_n, B_n, C_n, D_n)^{T}$, these
continuity constraints on the piecewise solution of 
Eq.\ (\ref{piece}) can be captured in a transfer matrix
\begin{equation}
\mathbf{v}_{n}=T_{n}\cdot\mathbf{v}_{n-1},
\end{equation}
where $T_{n}$ is given by
\begin{equation}
\left(
\begin{array}{cccc}
\frac{\beta_{n}}{f^{3/2}}S_{n}+K_{n} & \frac{\beta_{n}}{f^{3/2}}K_{n}-S_{n} & \frac{\beta_{n}}{f^{3/2}}\Delta x_{n} & \frac{\beta_{n}}{f^{3/2}} \\
S_{n} & K_{n} & 0 & 0 \\
-\frac{\beta_{n}}{f}S_{n} & -\frac{\beta_{n}}{f}K_{n} & 1-\frac{\beta_{n}}{f} \Delta x_{n} & -\frac{\beta_{n}}{f} \\
0 & 0 & \Delta x_{n} & 1
\end{array}
\right) 
\end{equation}
and

\begin{eqnarray}
\Delta x_{n} & \equiv & x_{n}-x_{n-1} \nonumber \\
S_{n} & \equiv & \sin[f^{1/2}(x_{n}-x_{n-1})],\nonumber \\
K_{n} & \equiv & \cos[f^{1/2}(x_{n}-x_{n-1})].\nonumber
\end{eqnarray}

At the two end-points at $x=0$, $\pi$, we have the boundary conditions
that $y$ and $y''$ vanish, which leads to the following four conditions
\begin{eqnarray}
B_0=D_0 &=&0 \label{bc1} \\
A_{N-1} S_N+B_{N-1} K_N&=&0 \label{bc2} \\
C_{N-1}(x_N-x_{N-1})+D_{N-1}&=&0. \label{bc3}
\end{eqnarray}
If we now define a matrix
\begin{equation}
\mathbf{R}=T_{N-1}T_{N-2}\ldots T_{2}T_{1}
\end{equation}
then Eqs.\ (\ref{bc1}-\ref{bc3}) lead to

\begin{equation}
\mathbf{M}\cdot
\left(
\begin{array}{c}
A_0
\\ 
C_0
\end{array}
\right)=0
\end{equation}
where 
\begin{equation}
\mathbf{M}\equiv
\left(
\begin{array}{cc}
R_{11} S_{N}+R_{21}K_{N} & R_{13} S_{N}+R_{23}K_{N}
\\
(\Delta x_{N})R_{31} +R_{41} &
(\Delta x_{N})R_{33} +R_{43}
\end{array}
\right)\label{M}
\end{equation}
For the beam to buckle, there needs to be non-zero
solutions for $A_0$ and/or $C_0$. Therefore, the
determinant of $\mathbf{M}$, which is a function
of $f$, must go to zero. The smallest $f$, $f_{\min}$,
at which $\det(\mathbf{M})=0$, gives the maximum 
compression tolerated by the beam and its support. 
The task, thus, is to find the set of $\{\beta_n \}$ 
and $\{ x_n \}$ which maximise $f_{\min}$.

\section{Equally spaced, equal springs}
Any definite choice of $\rho(x)$ provides a lower bound on the maximum
achievable value of $f_{\min}$, so before discussing the full numerical
optimization results on $\rho$, we consider here a simple choice of
$\rho$ which illuminates the physics.

\begin{figure}
\begin{center}
\includegraphics[width = 3.0in]{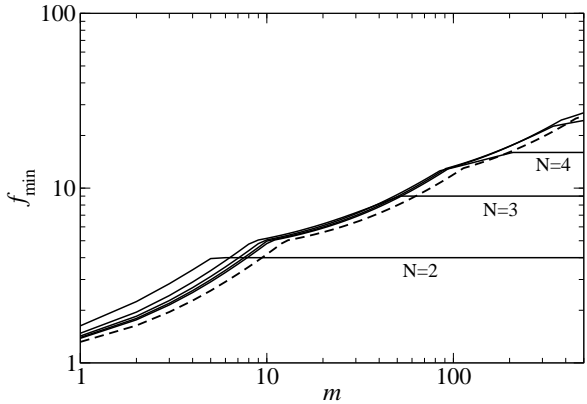}
\caption{\label{const_and_equal}
Value of $f_{\min}$ for $\rho$ constant (dashed line), and for equally spaced, equally
strong delta functions ($N$ is the number of intervals, so $N-1$
is the number of delta-functions).
}
\end{center}
\end{figure}

Suppose that $\rho$ consists of equally spaced, equally strong delta-functions:
\begin{equation}
\rho_{N}(x)=\sum_{n=1}^{N-1}\frac{m}{N-1}\delta(x-\pi/n).
\end{equation}
The value of $f_{\min}$ can be found by a straight-forward calculation
for each value of $m$, using the transfer matrix formulation above.
The results are plotted in figure \ref{const_and_equal}, and we see
that in general, it is better to concentrate the elastic support into
discrete delta functions, rather than having a uniform elastic support.
However, it is important to choose the appropriate number of delta functions:
if the number is too few, then there will always be a buckling mode 
with $f=N^2$ which threads through the comb of delta functions without
displacing them. However, apart from this constraint, it appears to be advantageous 
to choose a smaller value of $N$; in other words, to concentrate the support.

\section{Numerical optimization of the support}
\begin{figure}
\begin{center}
 \includegraphics[width=3.0in]{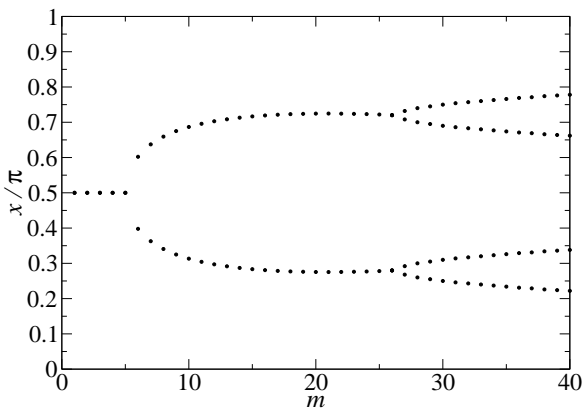}
\caption{\label{Res_bif}
Results of the restricted optimization, obtaining the set $\{x_n\}$ with constant $\beta_n=m/(N-1)$.}
\end{center}
\end{figure}
\begin{figure}
\begin{center}
 \includegraphics[width=3.0in]{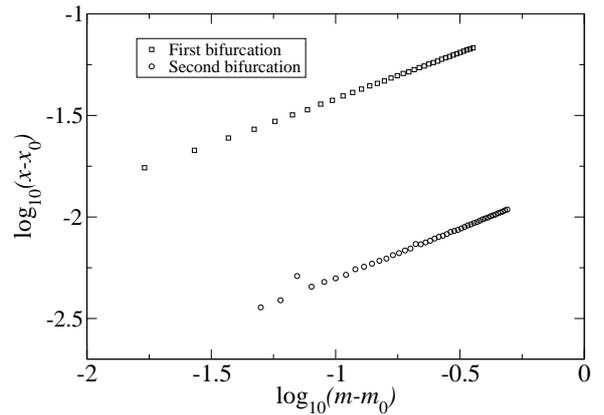}
\caption{\label{Exponents}
Showing the critical exponents for the first and second bifurcation in the restricted problem of $\beta_n = m/(N-1)$.}
\end{center}
\end{figure}

Before we look at the general optimization problem where we will 
seek the optimal set of $\{x_n\}$ and $\{\beta_n\}$ for a given cost, we investigate a 
simplified problem to give us further insight into the nature of the problem. We set
\begin{equation}
 \beta_n = \frac{m}{N-1} \quad \forall \quad n
\end{equation}
and then find the set $\{x_n\}$ which maximises $f_{min}$. The results obtained from an exhaustive 
search are shown in figure \ref{Res_bif}, where we find two bifurcation points in the range $0 < m\le40.$ The critical exponent of each has been obtained through simulation as, 
\begin{eqnarray}
 \alpha_1 &=& 0.5\pm 0.01, \\
 \alpha_2 &=& 0.49 \pm 0.03,
\end{eqnarray}
for the first and second bifurcation respectively. Figure \ref{Exponents} shows the data from which the exponents are taken, where values of $m_0$ and $x_0$ used are,
\begin{eqnarray}
 m_0 = 5.09,\; 26.99\\
 x_0 = 0.5, \; 0.281
\end{eqnarray}
for the first and second bifurcation respectively. The value of $x_0$ for the 
lower branching event at $m=26.99$ is related to the upper branch by
symmetry about the midpoint of the beam. As discussed previously, the optimal 
solution must split further at higher values of $m$. We hypothesize
that within this restricted problem these splits will take the form of bifurcations similar in nature to those 
found here. 

\begin{figure}
\begin{center}
\includegraphics[width=3.0in]{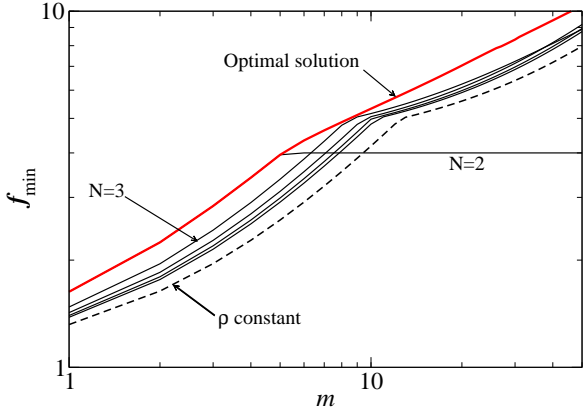}
\caption{\label{optimal_f}
Value of $f_{\min}$ for the optimal form of $\rho(x)$ and also for comparison
$\rho$ constant, and for equally spaced, equally
strong delta functions.
}
\end{center}
\end{figure}

Now we turn to the full optimization problem, where
the values $\{\beta_i\}$ as well as the positions $\{x_i\}$ of the
supports may vary. Using the transfer matrix formulation, we seek the optimal elastic support
consisting of delta functions. Figure \ref{optimal_f} shows the best solutions,
found from an exhaustive search of four delta functions ($N=5$), up to
$m=50$.
We see in figure 4 that there are two bifurcation events, and one coalescence of the
branches.  Because the optimal solution
cannot contain long intervals with no support (see section \ref{largem} below), we expect that if
 continued to larger values of $m$ and $N$,
a series of further bifurcation events would lead to a complex behaviour
which would eventually fill the interval with closely spaced delta
functions as $m\rightarrow\infty$. 

\begin{figure}
\begin{center}
\includegraphics[width=3.0in]{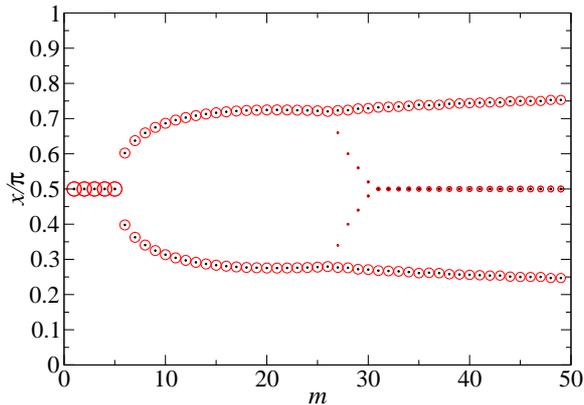}
\caption{\label{bif_diag}
Position of optimal springs as a function of $m$. 
The area of each circle is proportional to the strength $\beta_{i}$ of 
the relevant support, with the total area of all the circles at each value 
of $m$ chosen to be a constant, independent of $m$.
}
\end{center}
\end{figure}

\section{First branch point}

Numerical results (figure \ref{optimal_f}) indicate that although a single
delta function at $x=\pi/2$ is the optimal form for $\rho$ in the
limit $m\rightarrow 0$, at some point the optimal support bifurcates.

It is clear that this first bifurcation must happen at $f=4$, since this
represents the excitation of the first anti-symmetric buckling mode in
the unsupported beam, and the delta function at $x=\pi/2$ provides no
support against this mode. Although the value of $f$ at this first
branch point is clear, neither the value of $m$ at which it occurs,
nor the nature of the bifurcation are immediately obvious.

\begin{figure}
\begin{center}
\includegraphics[width=\columnwidth]{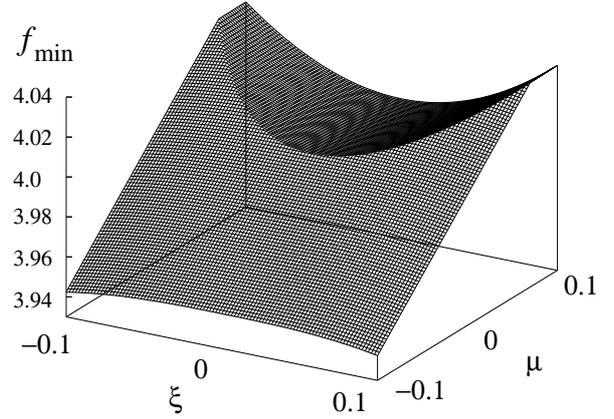}
\caption{\label{series}
Three dimensional plot of $f_{\min}$ as a function of the position
parameter $\xi$ and $\mu\equiv m-(16/\pi)$.
}
\end{center}
\end{figure}

In order to clarify the behaviour at this first branch point, we
perform a perturbation expansion: Let us suppose
that $N=3$ and
\begin{equation}
\rho(x)=\frac{m}{2}\delta\left(x-\frac{\pi}{2} +\xi\right)
+\frac{m}{2}\delta\left(x-\frac{\pi}{2} -\xi\right),
\end{equation}
where $\xi$ and $-\xi$ are clearly equivalent, and we will quote only the 
positive value later. Thus $\{ x_0, x_1, x_2, x_3 \}$ are given by 
$\{0,\pi/2-\xi,\pi/2+\xi,\pi \}$ and $\beta_1 = \beta_2 =m/2$.

We wish to evaluate the matrix $\mathbf{M}$ in Eq. (\ref{M}) and seek the 
smallest $f$ giving a zero determinant. On performing a series expansion
of the determinant for $f$ near $4$, we find that the critical value 
of $m$ is $16/\pi$. Furthermore, if we define small quantities
$\mu$ and $\xi$ through
\begin{eqnarray}
m=\frac{16}{\pi}+\mu\equiv \frac{16}{\pi}+\mu' \epsilon \\
\xi\equiv \xi' |\epsilon| 
\end{eqnarray}
where $\epsilon\ll 1$ and $\xi'$ and $\mu'$ are order $1$ quantities and
\begin{equation}
f=f(\xi,\mu),
\end{equation}
then we can perform a series expansion of $\det(\mathbf{M})$ 
in the neighbourhood of $\epsilon=0$, to obtain term by term a series 
expansion for $f$. We find that there are two solutions, $f_{+}$ and $f_{-}$,
which correspond to functions $y(x)$ symmetric and anti-symmetric about
$x=\pi/2$ respectively:
\begin{eqnarray}
f_{+}=\left[4+\frac{\pi}{6}\mu -\frac{\pi^2}{576}\mu^2
+\frac{\pi^3 (6-\pi^2)}{124416}\mu^3  \right.\nonumber \\\left.+\frac{\pi^4 (2\pi^2 -21)}{11943936}\mu^4 \right.\nonumber \\\left.+\frac{\pi^5\left(315-15\pi^2-\pi^4\right)}{4299816960}\mu^5+O(\mu^6)\right] \nonumber \\
+|\xi| \left[0 + O(\mu^5) \right] \nonumber \\
+\xi^2 \left[\frac{2\pi}{9}\mu+\frac{\pi^2}{72}\mu^2 -\frac{\pi^3\left(3+\pi^2\right)}{93312}\mu^3+O(\mu^4)\right] \nonumber \\
+|\xi^3 |\left[ -\frac{128}{9\pi}-\frac{40}{27}\mu -\frac{\pi\left(15-\pi^2\right)}{486}\mu^2+ O(\mu^3)\right]  \nonumber \\
+\xi^{4} \left[ 0+O(\mu^2)\right] +|\xi^5 |\left[
-\frac{1024}{135\pi}+O(\mu)\right] \label{f+}
\end{eqnarray}
\begin{eqnarray}
f_{-}=4+|\xi| \left[0 + O(\mu^5) \right]
+\xi^2 \left[\frac{32}{\pi^2}+\frac{2}{\pi}\mu +O(\mu^4)\right] \nonumber \\
+|\xi^3| \left[0 + O(\mu^3) \right]\nonumber\\
+\xi^4 \left[ -\frac{(128\pi^2 + 576)}{3\pi^4}-\frac{(8\pi^2 +72)}{3\pi^3}\mu + O(\mu^2)\right] \nonumber \\
+|\xi^5|\left[\frac{512}{3\pi^3}+O(\mu)\right]. \label{f-}
\end{eqnarray}

The final value for $f_{\min}$ in this neighbourhood is then 
$f_{\min}=\min(f_{+},f_{-})$.

The results are plotted in figure \ref{series}, and we see that the
behaviour of $f_{\min}$ around the bifurcation point is not analytic, since
the transition between the two branches $f_{+}$ and $f_{-}$ leads to
a discontinuity in the derivatives of $f_{\min}$. 
The maximal value of $f_{\min}$ (i.e the
optimum we are seeking), occurs for $\xi=0$ when $\mu<0$, and along the
locus $f_{+}=f_{-}$ when $\mu>0$. 

\begin{figure}
\begin{center}
\includegraphics[width=3.0in]{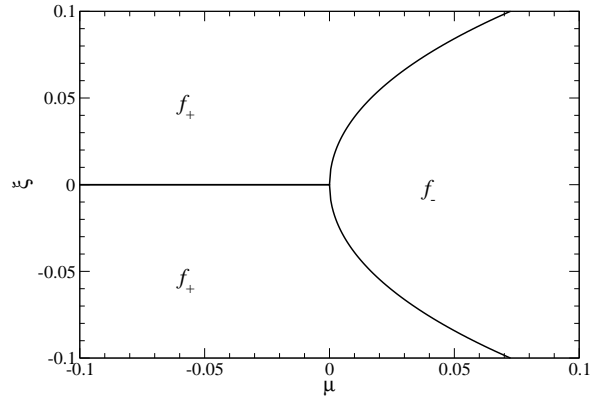}
\caption{\label{locus}
Curve shows the locus of optimal values for $\xi$ near the first
bifurcation point. This divides the $\xi-\mu$ plane into three regions,
in which $f_{\min}$ is given by either Eq. (\ref{f+}) or (\ref{f-})
as indicated.
}
\end{center}
\end{figure}

From Eqs. (\ref{f+}) and (\ref{f-}), this leads to the optimal value
of $\xi$ being
\begin{eqnarray}
\xi_{\rm opt} = 
\begin{cases}  \frac{\pi^{3/2}}{8\sqrt{3}}\mu^{1/2}-\frac{\pi^{4}}{864}\mu +O\left(\mu^{3/2}\right) \: & \text{if  $\:\mu \ge 0 $}
\\
0 \:& \text{if $\:\mu < 0$}
\end{cases}
\label{xi_opt}
\end{eqnarray}
This is shown in figure \ref{locus}, together with the regions of
the $\mu-\xi$ plane in which $f_{+}$ and $f_{-}$ apply.

\section{Limit of large support stiffness}\label{largem}
The results of our numerical optimisation suggests that the optimum 
support continues to take the form of a discrete set of delta-functions.
Here we investigate the possible form of the optimal support in the limit
of large $m$. 

As $m$ increases, the optimal distribution function
$\rho_{\rm opt}$ must become more evenly distributed over the interval.
To see in what sense this is true, we note that
the eigenvalue problem for buckling modes given by
Eq.\ (\ref{eq:eb}) can be derived from an energy approach:
Suppose that $z(x)$ is any deformation of the beam, then the energy of our
system is given \cite{Timoshenko} by
\begin{equation}\label{U}
U=\frac{1}{2}\int_{0}^{\pi}\left[\left(\frac{d^{2}z}{dx^{2}}\right)^{2}
-f\left(\frac{dz}{dx}\right)^{2}+\rho(x)z^{2}\right]{\rm d}x.
\end{equation}
Any deformation $z(x)$ which results in $U[z(x)]<0$ means that the 
beam will be energetically allowed to buckle under this deflection. Furthermore, the associated value of $f$ which just
destabilises the system against this deformation cannot be smaller
than the lowest buckling mode $f_{\min}$.

Consider therefore a particular choice for $z(x)$, namely 
\begin{equation}\label{z}
z(x)=\left\{
\begin{array}{ll}
0 & x\in(0,x_{1}) \\
\sin^{2}\left[\frac{\pi(x-x_{1})}{x_{2}-x_{1}}\right] & x\in(x_{1},x_{2}) \\
0 & x\in(x_{2},1) 
\end{array}\right. ,
\end{equation}
which vanishes everywhere except on the interval $\Omega=(x_{1},x_{2})$,
which is of length $\lambda\equiv x_{2}-x_{1}$.
Then Eq.\ (\ref{U}), together with the observation above about $f_{\min}$
leads to
\begin{equation}\label{U2}
f_{\rm min}[\rho(x)]\le\frac{2\pi^{2}}{\lambda^{2}}+\frac{\lambda}{\pi^{2}}
\int_{\Omega}\rho(x)\sin^{4}\left[\frac{\pi(x-x_{1})}{\lambda}
\right]{\rm d}x.
\end{equation}
Trivially, we note from the definition of $\rho_{\rm opt}$, that
\begin{equation}\label{op}
\forall \rho:f_{\min}[\rho(x)]\le f_{\min}[\rho_{\rm opt}(x)],
\end{equation}
so that from Eqs. (\ref{uni}), (\ref{U2}) and (\ref{op}), we finally arrive at
a condition for how evenly distributed $\rho_{\rm opt}$ must be for large $m$:
\begin{equation}\label{gaps}
\forall\Omega:
\int_{\Omega}\rho_{\rm opt}(x)\sin^{4}\left[\frac{\pi(x-x_{1})}{\lambda}
\right]{\rm d}x\ge
\frac{2\pi^{3/2}m^{1/2}}{\lambda}-\frac{2\pi^{4}}{\lambda^{3}}.
\end{equation}
A simple corollary of Eq.\ (\ref{gaps}) is that if $\rho_{\rm opt}$ is
zero on any interval $\Omega$ of length $\lambda$, then it must be
the case that
\begin{equation}\label{gaps2}
\lambda\le\pi^{5/4} m^{-1/4}.
\end{equation}
The scaling of this length with $m$ is the same as the effective buckling 
length of a uniformly supported beam discussed earlier.

\section{Discussion}
The optimal elastic support for our column appears to display complex 
behaviour: at small values of $m$ the support is
a single delta function, and even at large values of $m$, it appears to
be advantageous for $\rho(x)$ to be concentrated into discrete 
delta-functions rather than to be a smooth distribution.

Furthermore, the manner in which the system moves from a single to
multiple delta functions is not trivial, and appears to be through 
bifurcation events. In the full optimization problem we find that the
first bifurcation event occurs with critical exponent of one half. 
Inverting Eq.~(\ref{xi_opt}) and substituting it into either 
Eq.~(\ref{f+}) or (\ref{f-}) we find that,
\begin{equation}
 f_{\rm min} \approx 
\begin{cases}  4 + \frac{32}{\pi^2}\xi_{\text{opt}}^2 - 
\frac{64\left(2\pi^2-9\right)}{3\pi^4}\xi_{\rm opt}^4 \: & \text{if  $\:\mu \ge 0 $}
\\
4+\frac{\pi}{6}\mu -\frac{\pi^2}{576}\mu^2
   \:& \text{if $\:\mu < 0$.}
\end{cases}
\end{equation}
while to leading order,
\begin{equation}\label{xiopt}
\xi_{\rm opt} = 
\begin{cases}  \frac{\pi^{3/2}}{8\sqrt{3}}\mu^{1/2} \: & \text{if  $\:\mu \ge 0 $}
\\
0\:& \text{if $\:\mu < 0$.}
\end{cases}
\end{equation}
In this form, the mathematical similarities to Landau theory
of second order phase transitions become apparent, with
$\xi_{\rm opt}$ playing the role of the order parameter, $\mu$ the 
reduced temperature and $-f_{\min}$ the free energy to be minimized.

However, there is an important difference.
In Landau theory of second order phase transitions,
the free energy $F_{\rm lan}$ is assumed to be a power series expansion
in the order parameter $\psi$ with leading odd terms missing:
\begin{equation}
F_{\rm lan}=F_0 + a_2 \psi^2 +a_4 \psi^4 + ...
\end{equation}
where $a_2 \propto (T-T_c)$, the reduced temperature.
In our case, the buckling force $f$ has to be first optimised
for even and odd buckling. Thus $-f_{\min}$ (which is the analogue of $F_{\rm lan}$) is
a minimum over two intersecting surfaces (figure \ref{series})
and so non-analytic at the point of bifurcation.

Nevertheless, the mathematical form of the solution
in Eq.~(\ref{xiopt}) is the same, including the critical exponent.
Furthermore, our numerical results show that, for the equal 
support case, the critical exponent $\alpha$ is preserved 
for the next bifurcation, suggesting that the
nature of subsequent bifurcations will also remain the same.

The details of the behaviour for larger values of $m$ is as yet unclear:
we speculate that there will be a cascade of bifurcations,
as seen in the limit set of certain iterated maps
\cite{Feigenbaum}; it remains an open question whether there is an accumulation
point leading to potential chaotic behaviour.

Further investigation of this regime may shed light on 
structural characteristics required to protect more complex engineering
structures against buckling instabilities.

\section{Acknowledgements}
The authors wish to thank Edwin Griffiths for useful discussions. The figures
were prepared with the aid of `Grace' 
({\tt plasma-gate.weizmann.ac.il/Grace}), 
`gnuplot' ({\tt http://www.gnuplot.info}) 
and `xfig' ({\tt www.xfig.org}). Series expansions were derived
with the aid of `Maxima' ({\tt maxima.sourceforge.net}).

\end{document}